\pdfoutput=1
\documentclass[aps,prl,twocolumn,superscriptaddress]{revtex4-2}
\usepackage[colorlinks=true,citecolor=blue,urlcolor=blue,linkcolor=blue,pdfstartview=FitH,bookmarksopen]{hyperref}
\usepackage{amsmath,amssymb}
\usepackage{bm}
\usepackage{comment}
\usepackage{graphicx,xcolor}
\newcommand\<\langle
\renewcommand\>\rangle
\newcommand\+\dagger
\newcommand\p{\bm{p}}
\newcommand\q{\bm{q}}
\newcommand\x{\bm{x}}

\newcommand\grad{\bm{\nabla}}
\renewcommand\d{\partial}
\renewcommand\Re{\mathop{\text{Re}}}
\renewcommand\Im{\mathop{\text{Im}}}

\newcommand\rxi{ R}

\begin{document}

\title{Fate of Multiparticle Resonances: From $Q$-Balls to
$^3$He Droplets}

\author{Dam Thanh Son}
\affiliation{Kadanoff Center for Theoretical Physics, University of Chicago, Chicago, Illinois 60637, USA}
\author{Mikhail Stephanov}
\affiliation{Department of Physics, University of Illinois, Chicago, Illinois 60607, USA}
\author{Ho-Ung Yee}
\affiliation{Department of Physics, University of Illinois, Chicago, Illinois 60607, USA}

\begin{abstract}

We consider a system of $N$ nonrelativistic particles which form a
near-threshold resonance. Assuming no subset of these particles can
form a bound state, the resonance can only decay through an
``explosion'' into $N$ particles.  We show that the decay width of the
resonance scales as $E^{\Delta-5/2}$ in the limit when the energy $E$
of the resonance goes to zero, where $\Delta$ is the ground state
energy of a system of $N$ particles in a spherical harmonic trap with
unit frequency.  The formula remains valid when some pairs of final
particles have zero-energy $s$-wave resonance, but the
Efimov effect is not present.  In the limit of large $N$, we show that
the final particles follow a Maxwell-Boltzmann distribution if they
are bosons, and a semicircle-like law if they are fermions.  We argue
that metastable $^3$He droplets exist with the lifetime varying over
many orders of magnitudes ranging from a fraction of a nanosecond to
values greatly exceeding the age of the Universe.

\end{abstract}

\maketitle

\emph{Introduction.}---The existence of ``Borromean'' states---bound
states of three particles, of which no pair is capable of forming a
bound state---and their generalization to more than three particles
(the ``Brunnian'' states) are of great interest in nuclear and atomic
physics~\cite{Zhukov:1993aw}.  In particular, much effort has been
dedicated to the search for universal properties of these systems.
The limit of zero-range interaction, where the Efimov effect is at
play, has received most attention; it has been shown that three- and
four-particle Efimov states have universal
properties~\cite{Hammer:2006ct,Stecher:2008}.  For more than four
particles, our knowledge is much more limited (see
Ref.~\cite{Naidon:2016dpf} for a review and further references).

In this Letter we are concerned not with many-particle {\em bound} states,
but with many-particle \emph{resonances}~\cite{Kukulin1989}.  We
address here a sharp question concerning the width of
multiparticle resonances in the near-threshold regime: what is the
behavior of the width of a resonance when its energy crosses zero,
i.e., when the resonance is just about to become a bound state
(for example, when a parameter
characterizing the interaction is varied)?  Our result shows that the
asymptotic behavior of the decay width is universal,
\begin{equation}\label{result}
  \Gamma(E) \sim E^{\Delta-5/2},
\end{equation}
where $\Delta$ is the ground state energy of a system of $N$
``surrogate'' particles in a spherical harmonic potential with unit
oscillator frequency for all particles, provided that
$\Delta>\frac72$.  The ``surrogate'' particles have the same
properties as the particles that make up the resonance (mass, spin,
statistics).  The interaction between the surrogate particles is
turned off, unless when a pair of the original particles have infinite
$s$-wave scattering length, in which case the corresponding surrogate
particles have  zero-range,
infinite scattering length (i.e., unitarity) interaction.
The significance of $\Delta$ is that it is
the conformal dimension~\cite{Nishida:2007pj} of the
lowest-dimensional operator that creates the resonance from the
vacuum.

Our result can be applied to various physical contexts where
multiparticle resonances appear.  For bosons interacting through a
potential of the Lennard-Jones type, in a certain range of the de Boer
parameter, bound clusters exist but only when the number of particles
exceeds a critical value~\cite{Hanna:2006,Zwerger2019}.  Metastable
droplets then should appear at particle numbers slightly smaller than
the critical value.  Another example is droplets of $^3$He atoms.  It
is known that $^3$He atoms form a bound droplet only when there is
sufficient number of them.  The minimal number of atoms in a bound
$^3$He droplet, $N_0$, has been estimated to be between 20 and
40~\cite{Pandharipande:1986,Barranco:1997,Guardiola:2000zz,Sola:2006,Sola:2007}.
Metastable $^3$He droplets can then appear when the number of atoms
$N$ is slightly smaller than $N_0$, e.g., for $N=N_0-1$.  We are not
aware of any previous estimate of the lifetimes of such metastable
nanodroplets of $^3$He.  Quantum droplets may exist in weakly-coupled
bosonic mixtures~\cite{Petrov:2015}. In relativistic quantum field
theory, a scalar quantum field theory that supports
$Q$-balls~\cite{Lee:1991ax,Coleman:1985ki} also allows for metastable
$Q$-balls~\cite{Levkov:2017paj}.

\emph{Previously known results.}---Before presenting arguments leading
to Eq.~(\ref{result}), let us check that it is consistent with all
previously known results.  For a two-body resonance with angular
momentum $\ell$, the energy of the surrogate system in the spherical
harmonic trap with unit frequency is $3+\ell$, and Eq.~(\ref{result})
then reproduces the known result $\Gamma\sim E^{\ell+1/2}$ for
$\ell\geq1$.  For three bosons with no resonant interaction, the
ground-state energy of the surrogate system is $\frac92$, giving
rise to the behavior $\Gamma\sim E^2$ previously found in
Ref.~\cite{Matsuyama:1991bm}.  When two of the three particles have
infinite scattering length, the resonance interaction reduces the
ground state energy of the surrogate system by $1$.  Now
Eq.~(\ref{result}) yields $\Gamma/E\sim E^0$, but as we will see, a
more careful analysis reveals that there is a logarithmic
modification which makes $\Gamma/E$ decrease logarithmically as
$E\to0$, as first found in Ref.~\cite{Konishi:2017lbg}.

\emph{New results.}---We can now read out the behavior of $\Gamma$ for some cases
which have not been solved before.  The most nontrivial predictions
involve spin-$1/2$ fermions at unitarity.  For a resonance formed from
two spin-up and one spin-down fermions of the same mass, with infinite
$s$-wave scattering length between two fermions of different spins
(an approximation for neutrons), the
ground state in a harmonic trap has energy $\Delta=4.27272$ for
$\ell=1$ and $\Delta=4.66622$ for $\ell=0$~\cite{Tan:2004,Werner:2006zz}.  The width of a
near-threshold resonance then behaves as
\begin{equation}
  \Gamma(E) \sim \begin{cases}
    E^{1.773}, & \ell=1,\\
    E^{2.166}, & \ell=0.
  \end{cases}
\end{equation}
In the case of neutrons, this behavior should hold for the trineutron
resonance if such a resonance exists with energy between
$E_a=\hbar^2/m_na^2\approx0.1$~MeV and $E_{r_0}=\hbar^2/m_nr_0^2\approx5$~MeV where $m_n$ is the
neutron mass, $a$ and $r_0$ the scattering length and effective range
of the $nn$ scattering, respectively.
If the energy of the resonance is less than
$E_a$, the behavior of $\Gamma$ is dictated by the ground-state energy
of three \emph{free} particles in the harmonic potential, which is
$\frac{11}2$ for $\ell=1$ and $\frac{13}2$ for $\ell=0$.  We find
$\Gamma\sim E^3$ and $\Gamma\sim E^4$ for these two cases.  A
near-threshold three-neutron resonance does not seem to exist the real
world~\cite{Kezerashvili:2016ucn,Marques:2021mqf}, but in model
calculations it appears when a sufficiently strong three-body
attraction is added to the forces between
neutrons~\cite{Lazauskas:2005fy}.  These behaviors are similar to the
``unnuclear'' behavior of nuclear reactions with emission of a few
neutrons~\cite{Hammer:2021zxb}.

For a four-neutron resonance (which appears if sufficiently strong
four-body attraction is added~\cite{Lazauskas:2005ig}) with energy in
the regime $E_a \ll E \ll E_{r_0}$, the behavior of the width is
controlled by the energy of the ground state of four unitary
fermions in a spherical harmonic trap, which was numerically
determined to be
$\Delta\approx5.0$ \cite{Chang:2007zzd,vonStecher:2007zz,Alhassid:2007cda,vonStecher:2009mu,Rotureau:2010uz,Endres:2011er,Rotureau:2013exe},
so $\Gamma\sim E^{2.5}$.  At energies much lower than $E_a$
the behavior becomes $E^{11/2}$.

\begin{figure}[ht]
\begin{center}
\includegraphics[width=0.5\textwidth]{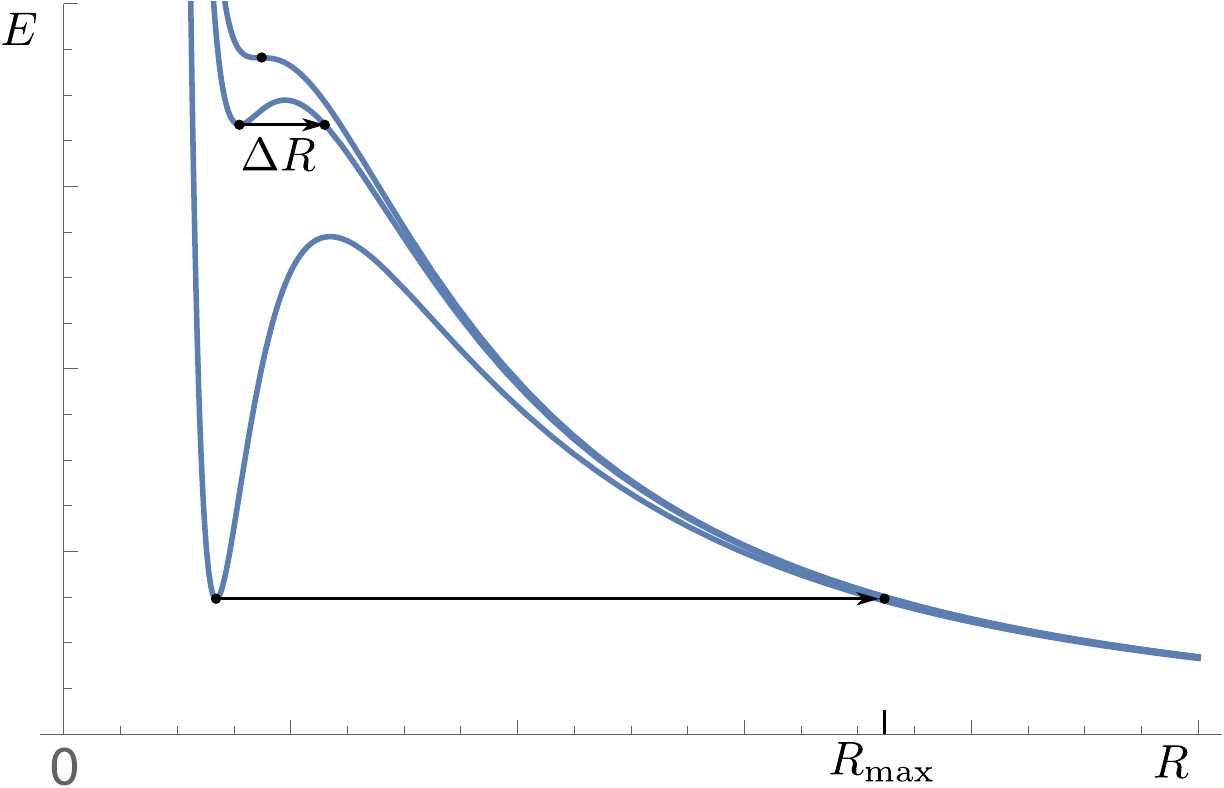}
  \caption{The droplet's energy as function of its size $\rxi$ for three values
    of $N$.  The upper curve corresponds to $N=N_1$, the middle curve
    to $N=N_1(1+\epsilon)$ where $0<\epsilon\ll1$, and the lowest
    curve to $N=N_0(1-\epsilon)$.}
\label{fig:var_en}
\end{center}
\end{figure}

\emph{Weakly coupled bosonic droplets.}---To gain intuition on the
problem, let us first consider metastable droplets of
bosons with small negative scattering length and effective three-body
repulsion~\cite{Bulgac:2001wc,Zwerger2019,Son:2020yiy}.  The Hamiltonian
of the model reads
\begin{equation}
  H[\psi] = \int\!d\x \left( \frac{|\nabla\psi|^2}2 - \frac{g}4|\psi|^4
      + \frac G6 |\psi|^6\right).
\end{equation}
(Here we set $\hbar=m=1$.)
When $G/g^4\gg1$, the droplets contain a large number of bosons
(which are the nonrelativistic version of $Q$-balls)
and can be found by minimizing the functional $H[\psi]$ at fixed
number of particles.  Solving the problem numerically, we find that
$H$ has a local minimum with positive energy for $N_1<N<N_0$ where
\begin{equation}
  N_1 \approx 189.4 \frac{G^{1/2}}{g^2} \,, \quad
  N_0 \approx 240.4 \frac{G^{1/2}}{g^2} \,.
\end{equation}
The parametric dependence of $N_0$ on $g$ and $G$ has been previously
predicted in Ref.~\cite{Zwerger2019}.
One can visualize the metastable droplet as the local minimum of the
function that gives the energy as a function of the size of the
droplet (Fig.~\ref{fig:var_en}).

The decay of a metastable droplet is described by an instanton, i.e.,
a solution to the equation of motion in Euclidean time. The instanton
can be found mostly analytically for $N$ near $N_1$ or $N_0$.
For $N=N_1$, there is a flat
direction in the functional space of the droplet density profiles. For
$N>N_1$, moving along this direction towards larger droplet size $\rxi$
one encounters a potential barrier, as shown in
Fig.~\ref{fig:var_en}. For small $N-N_1$ the width $\Delta\rxi$ of the
barrier shrinks as $\Delta\rxi\sim (N-N_1)^{1/2}$ and vanishes at
$N=N_1$. In this regime one can calculate the tunneling
amplitude using WKB approximation for the effective action in
collective coordinate $\rxi$ with potential
$U(\rxi)\sim N\left[\Delta\rxi(\rxi-\rxi_{\rm eq})^2-(\rxi-\rxi_{\rm
    eq})^3\right]$ which has a metastable minimum at
$\rxi=\rxi_{\rm eq}$ and a point of exit from the ``tunnel'' at
$\rxi=\rxi_{\rm eq}+\Delta\rxi$ shown in Fig.~\ref{fig:var_en}. The
imaginary action for classically forbidden tunneling is given by
$S_I\sim \int_{\rxi_{\rm eq}}^{\rxi_{\rm
    eq}+\Delta\rxi}\!d\rxi\sqrt{NU(\rxi)}\sim N(\Delta\rxi)^{5/2}$ resulting
in the exponentially suppressed decay rate
$\Gamma\sim\omega\sqrt{S_I}\exp(-2S_I)$, where
$\omega\sim\sqrt{\Delta\rxi}$ is the frequency of the harmonic motion
near the local minimum of $U(\rxi)$, or
\begin{equation}
  \Gamma = \frac{c_2\sqrt{N_1}}{\sqrt G}\left(\frac{N{-}N_1}{N_1}\right)^{7/8}
  \exp\Biggl[ - c_1 N_1 \biggl(\frac{N{-}N_1}{N_1}\biggr)^{5/4}\Biggr],
\end{equation}
where $c_1\approx1.58$ and $c_2\approx 0.570$ \cite{to-appear}.

At the other end of the window of metastability, near $N=N_0$, the
system has to tunnel in Euclidean time to a droplet of a very large
size before it can expand classically in real time.  The energy of a
cloud of $N$ bosons with size $R$ is $E\sim N/(mR^2)$, so coming out from
under the barrier, the cloud of particles has size
\begin{equation}
  R_\text{max} \sim \sqrt{\frac N{mE}}\,,
\end{equation}
which diverges as $E\to0$.  In contrast, the size of the system at the
beginning of the tunneling process, $R_\text{min}$ remains finite as
$E\to0$ (see Fig.~\ref{fig:var_en}).  Most of the tunneling thus
occurs in the regime where the inter-particle interaction can be
neglected.  Since the potential energy behaves like $1/R^2$, the WKB
exponent is proportional to $\ln(R_\text{max}/R_\text{min})$.  To find
the exact numerical coefficient, we need to solve the Euclidean
equations of motion.  Writing $\psi=f e^{i\theta}$, in Euclidean time
$\tau=it$ and $\varphi=-i\theta$, the Euclidean action becomes
\begin{equation}\label{eq:S_E}
  S_E = \int\!d\tau\, d\x \left( - f^2 \d_\tau\varphi
  -\frac{f^2}2 (\grad\!\varphi)^2 + \frac{(\grad f)^2}{2}
  \right).
\end{equation}
One can check that the following configuration is a solution to the
Euclidean field equations with $E\to0$
\begin{align}\label{Euclidean-sol}
  f & = \frac{\sqrt N}{(2\pi\tau)^{3/4}}
  \exp\left(-\frac{r^2}{4\tau} \right) ,\\\label{eq:phi}
  \varphi &= \frac{r^2}{4\tau} + \frac34 \ln\tau .
\end{align}
Note that Eq.~(\ref{eq:phi}) corresponds to a ``Hubble expansion,''
$\grad\varphi= \bm r/(2\tau)$.  The solution applies in the intermediate
regime when the size of the droplet $\tau^{1/2}$ is larger than the
original size, but much smaller than the droplet size when it exits
from under the barrier.  Evaluating the Euclidean action of the
solution (\ref{Euclidean-sol}) we find
\begin{equation}
  S_E = \frac{3N}2 \ln \frac{R_\text{max}}{R_\text{min}} \,.
\end{equation}
The decay rate is $\Gamma\sim e^{-2S_E}\sim R_\text{max}^{-3N}$.
Since $R_\text{max}\sim E^{-1/2}$, we find that $\Gamma\sim E^{3N/2}$.
At large $N$, where the semiclassical instanton calculation applies,
the resonance is narrow, i.e. $\Gamma\ll E$.

\emph{Field theory approach.}---The above approach is not applicable
when the number of particles in the droplet is small, or when
they are fermions.  In these cases one can still find the behavior of
the width of the resonance when its energy is small using a low-energy
effective field theory.  Let $\Psi$ be the field describing the
resonance, and $\psi_a$ are the particles that constitute this
resonance (which may belong to different species $a$).  The
effective field theory describing the system is
\begin{multline}
  \mathcal L = \Psi^\+ \left( i\d_t + \frac{\nabla^2}{2m_\Psi}\right) \Psi
  + L[\psi_a]
  + \mu_0\Psi^\+\Psi\\
  + g (O^\dagger\Psi + \Psi^\dagger O) ,
\end{multline}
where $L[\psi_a]$ is the Lagrangian of nonrelativistic conformal field
theory (NRCFT)~\cite{Nishida:2007pj} of the $\psi$-particles (the
simplest version of a NRCFT is a free field theory), $O$ is a
(composite) operator with conformal dimension $\Delta$ in the theory
described by $L[\psi_a],$ $g$ and $\mu_0$ are some parameters.  The
simplest example of $O$ is $O=\psi^N$ in the case where $\psi$ is a
boson field, where $\Delta=\frac32N$.  The coupling can be considered
point-like if the excitation energy of the droplet is larger than the
typical energy of the final particles, which is the case when the
resonance is near threshold.  The field theory is assumed to have an
ultraviolet cutoff at momentum scale $\Lambda$ (energy scale
$\Lambda^2$).

The self-energy of $\Psi$ obtained by integrating out $\psi$ is
\begin{equation}
  \Sigma(\omega,\q) = -i g^2 \< O O^\+\>_{\omega,\q} \,.
\end{equation}
Galilean invariance implies that $\Sigma$ is a function of
$E=\omega-q^2/(2m_\Psi)$.  The
correlator of $O$, in general, contains ultraviolet divergences which
are regularized by the cutoff $\Lambda$.  These UV divergences
contribute to the real (but not the imaginary) part of $\Sigma$.

We expand $\Sigma$ in powers of $E$, keeping only the first two terms in the
real part and the leading term in the imaginary part. When $\Delta>\frac72$, the
first two terms in the real part have power-law divergences, and the
result reads
\begin{multline}\label{Sigma}
  \Sigma(\omega,\q) =
  - g^2 \bigl[
  a_0 \Lambda^{2\Delta-5}
  + a_1 \Lambda^{2\Delta-7} E\\
  + i b_0 E^{\Delta-5/2}\theta(E) \bigr] ,
\end{multline}
where $a_0$, $a_1$, $b_0$ are some numbers.  The existence of a
low-energy resonance means that
$\mu=\mu_0+g^2a_0\Lambda^{2\Delta-5}$ is fine-tuned to an
unnaturally small value; the $a_1$ term leads to a wavefunction
renormalization for $\Psi$: $Z^{-1}=1+g^2 a_1\Lambda^{2\Delta-7}$.
The propagator of $\Psi$ is now
\begin{equation}
  \< \Psi \Psi^\+\> \sim \left[
  Z^{-1} E + \mu + ig^2b_0 E^{\Delta-5/2}\theta(E)
  \right]^{-1} .
\end{equation}
When $\mu$ is small, the propagator's pole is located at $E_*=\Re
E_*+i\Im E_*$, where $\Re E_*=-Z\mu$, and
\begin{equation}
   \Im E_* = - g^2 b_0 Z (\Re E_*)^{\Delta-5/2} ,
\end{equation}
which goes to zero faster than $\Re E$ for $\Delta>\frac72$.  For
example, for a decay resonance consisting of three bosons in $s$-wave,
$\Delta=\frac92$ and $\Im E\sim (\Re E)^2$, as found in
Ref.~\cite{Matsuyama:1991bm} using a different method.

Consider now the case $\Delta=7/2$.  This case corresponds to the
resonance consisting of two particles in $s$-wave resonance and a
third particle of a different type that does not interact resonantly
with any of the first two.  The resonant pair is described by a
``dimer'' field with dimension~2~\cite{Nishida:2007pj}, and the third
particle by a free field of dimension~$\frac32$, so the total
dimension of $O$ is $2+\frac32=\frac72$.
In this case
\begin{equation}
  \Sigma(\omega,\q) = -\Bigl[ a_0 \Lambda^2 + a_1 E\ln\frac{\Lambda^2}{|E|} +
  i \pi a_1 E\,\theta(E) \Bigr] ,
\end{equation}
and redoing the analysis one sees that
the ratio between the imaginary and real parts of the position of the
pole decreases logarithmically with the energy.  This was previously
found in Ref.~\cite{Konishi:2017lbg}.

We now rederive Eq.~(\ref{result}) using a different method, which
allows us to gain additional intuition for the behavior, and will also
give us additional information about the decay.  In particular, for a
resonance of $N\gg1$ particles we will find the momentum distribution
of the final particles.

\emph{Decay as tunneling through centrifugal barrier.}---The
suppression of the decay rate as $E\to0$ can be interpreted as the
result of tunneling under a barrier.  Instead of the position of $N$
particles one can introduce the center of mass coordinate, one
hyperradius, and $N-2$ hyperangles.  Factoring out the center-of-mass
motion, the Schr\"odinger equation with no interaction
can then be written as
\begin{equation}
  \frac{\d^2\psi}{\d R^2} + \frac{3N{-}4}R
  \frac{\d\psi}{\d R} + \frac{\Delta_\Omega\psi}{R^2}
  = 0,
\end{equation}
where $\Delta_\Omega$ is the Laplacian operator in hyperangles.  For
bosons the lowest eigenvalue of $-\Delta_\Omega$ is 0, which
corresponds to the solution $\psi\sim R^{-3N+5}$.  The decay rate can
be obtained by evaluating the probability flux at $R=R_\text{max}$:
$R^{3N-4}\psi^*\d_R\psi\sim R^{-(3N-5)}$.  For $R_\text{max}\sim
E^{-1/2}$, this implies $\Gamma\sim E^{(3N-5)/2}$~\footnote{This is
essentially the
$N$-body generalization~\cite{Sadeghpour_2000} of the Wigner
threshold law.}.

For fermions or particles interacting with $s$-wave resonance in general, the 
picture of the decay as tunneling under a $1/R^2$ barrier is still
valid~\cite{to-appear}.  From the mapping between the dimension of primary operator and the energy in a harmonic trap, the coefficient of the $1/R^2$ potential is determined to be $(\Delta-2)(\Delta-3)/2$.
The same discussion as in the bosonic case then gives the decay rate scaling as $E^{\Delta-5/2}$ in agreement with the field theory approach. 
One can also
make use of the SO(2,1) symmetry of NRCFT to arrive at the same conclusion.

\emph{Momentum distribution of final particles.}---We now ask the
following question: What is the momentum distribution
of the final decay products of a metastable droplet of $N$ particles,
where $N\gg1$?

For bosons, this distribution can be derived from the following
argument: the amplitude of the decay of one resonance into $N$ bosons $\Psi\to
N\psi$ should be independent of the momenta of final particles when
the latter are small.  This implies that that the distribution of
final particles over momentum is the same as in a microcanonical
ensemble of $N$ bosons where the energy is fixed to the energy of the
resonance, $E$, and the momentum to 0.  In the limit of a large number
of particles, the ensemble is equivalent to the canonical ensemble,
hence the final particles should follow the Maxwell-Boltzmann
distribution,
\begin{equation}
  \frac{dN}{d\p} \sim \exp \left( - \frac{\bm p^2}{2mT_\text{eff}} \right),
\end{equation}
with the effective temperature determined by the total energy, i.e.,
$T_\textrm{eff}=\frac23 E/N$.
The same result follows from the Gaussian shape of the droplet's
wave function with width $\rxi_{\rm max}=\sqrt{3N/4E}$
at the end of the tunneling
described by the Euclidean action in Eq.~(\ref{eq:S_E}).

Now consider a resonance consisting of $N$ fermions.  For simplicity, let us consider spinless fermions.
The vertex describing the decay of the resonance is now
\begin{equation}\label{f-vertex}
  \mathcal L_\text{int} = g \Psi^\+ \psi
  \d_x\psi \d_y\psi \d_z\psi
    \d_x^2\psi \d_x\d_y \psi\cdots
\end{equation}
The probability distribution function of the particles over their
momenta is then
\begin{multline}\label{pdf}
  \rho(\p_1,\p_2,\ldots \p_n) \sim
  \begin{vmatrix}
  P_1(\p_1) & P_1(\p_2) & \ldots & P_1(\p_N)\\
  P_2(\p_1) & P_2(\p_2) & \ldots & P_2(\p_N)\\
  \ldots & \ldots & \ldots & \ldots \\
  P_N(\p_1) & P_N(\p_2) & \ldots & P_N(\p_N)
  \end{vmatrix}^2\\
  \times \delta\Bigl(\sum_a \p_a\Bigr) \delta \Bigl( \sum_a \frac{\bm p_a^2}{2m}-E\Bigr),
\end{multline}
where $P_1(\p)=1$, $P_2(\p)=p_x$,
$P_3(\p)=p_y$,
$\ldots$ are monomials of $\p$, each corresponding to a factor
in the vertex~(\ref{f-vertex}).  In the limit of large $N$ we can
replace the delta functions by the exponential factor $\exp
[-\beta\sum_a \bm p_a^2/2m ]$.  It can be seen that the probability
distribution function~(\ref{pdf}) is the square of the wavefunction of
a ground state of $N$ fermions in a harmonic potential with a suitable
frequency.  The distribution of final particles over momentum can be
obtained through the Thomas-Fermi approximation of particles in a
harmonic trap.  The result is
(the formula also works for spin-$\frac12$ fermions)
\begin{equation}
   \frac{dN}{d\p} \sim \left(p_\text{max}^2-\bm p^2\right)^{3/2} ,
\end{equation}
where $p_\text{max}^2=16mE/(3N)$.

\emph{Lifetime of metastable $^3$\textup{He} droplets}.---According to Monte
Carlo calculations, a cluster of $N$ $^3$He atoms becomes bound at
some $N=N_0$ between 20 and
40~\cite{Pandharipande:1986,Barranco:1997,Guardiola:2000zz,Sola:2006,Sola:2007}.
Thus there must be a range of $N$, $N_1< N< N_0$, where the droplet
has positive energy and is metastable.  For $N$ slightly smaller than
$N_0$, the energy of the droplet is small, so its lifetime must be
large.  For example for $N=20$ the energy per atom in the droplet was
estimated to be about
0.2~K~\cite{Pandharipande:1986,Barranco:1997,Guardiola:2000zz}, much
smaller than the binding energy per particle of infinite $^3$He liquid
($2.4$~K).  For $N$ just below $N_0$ the energy of the
droplet is even smaller, typically less than 1~K for the whole
droplet.

To estimate the lifetime of a metastable $^3$He droplet, we note that
the energy of noninteracting $N$ spin-$\frac12$ particles in a
harmonic trap of unit frequency is
\begin{equation}
  \Delta = 60 + \frac92 (N-20)
\end{equation}
for $20\le N\le40$.  Taking $N_0$ to be the smallest number quoted in
the literature, $N_0=29$, we consider the metastable droplet with
$N_0-1=28$ atoms, $\Delta=96$.  This leads to a huge power in the
dependence of the width on the energy: $\Gamma\sim (E/E_0)^{93.5}$.

$E_0$ can be estimated to be the kinetic energy of a free Fermi gas of
$N$ particles, confined by a harmonic potential with frequency chosen
so that the rms size of that cloud of particles is equal to
the rms size of the metastable droplet.  In the Thomas-Fermi
approximation, the kinetic energy of a cloud of $N$
particles in a harmonic potential is related to its rms size
$\<r^2\>^{1/2}$ by
\begin{equation}
  E_0 = \frac{3^{8/3}}{32} \frac{N^{5/3}}{m\<r^2\>} \,. 
\end{equation}
For $N=28$ and $\<r^2\>^{1/2}\sim 8$~\AA~\cite{Pandharipande:1986}, we
find $E_0\sim 40$~K.  For $E\sim 1$~K the suppression factor
$(E/E_0)^{93.5}$ becomes $10^{-150}$.  Even with the large uncertainty
in the estimate, it is obvious  a $^3$He droplet
containing one or a few particles less than the smallest stable
droplet should live longer than the age of the Universe.  These droplets,
though having positive energy (relative to the free atoms), are
essentially stable.

As the number of atoms in the droplet decreases, the lifetime becomes
shorter and, at some value must become comparable to
$\hbar/\text{K}\sim 10^{-11}$~s.  By varying number of particles the
lifetime of the $^3$He droplet can vary from a fraction of a
nanosecond to values much larger than the age of the Universe.
Unfortunately at this moment we have no method that can tell us
reliably the lifetime of a $^3$He droplet with a given $N$, nor can we
say for which numbers of atoms the lifetime of the droplet may be in
the experimentally interesting range.

\emph{Conclusion.}---In this Letter we have shown that the
near-threshold $N$-body resonances have certain universal properties.
The lifetime of the resonance scales with energy with a universal
exponent.  The momentum distribution of the final decay products is also
universal.

We have shown that in nature, metastable $^3$He droplets exist in a range of
sizes.  It would be useful to quantitatively determine that range
and the lifetime of droplets with size therein.  While the energetics of small $^3$He
droplets can be determined using various numerical
methods, the study of the lifetime will likely require the development of
new approaches.  We hope that metastable droplets of
ultracold atoms that decay into individual atoms can also be created and
studied in the laboratory.

The authors thank Gia Dvali, Jeff Harvey, Andrey Shirokov, Peter
Tinyakov, and Wilhelm Zwerger for valuable comments.  This work is supported, in part, by
the U.S. Department of Energy, Office of Science, Office of Nuclear
Physics, within the framework of the BEST Topical Collaboration and
grant No.\ DE-FG0201ER41195, by the U.S.\ DOE grant No.\
DE-FG02-13ER41958, by a Simons Investigator grant and by the Simons
Collaboration on Ultra-Quantum Matter, which is a grant from the
Simons Foundation (651440, DTS).

\bibliography{droplet}

\end{document}